\documentstyle[prb,preprint,tighten,eqsecnum,aps,epsf]{revtex}
\begin{document}
\draft
\title{The effects of electron-electron interactions on the spin 
transport dynamics of a two-dimensional electron gas}
\author{Yutaka~Takahashi\cite{email}}
\address{Department of Electrical Engineering,
Yamagata University, Yonezawa, Yamagata 992-8510, Japan}

\author{Kosuke~Shizume}
\address{University of Library and Information Science, 1-2 Kasuga, Tsukuba,
305-8550, Japan}

\author{Naoto Masuhara}
\address{Department of Physics, University of Florida, Gainesville, Florida,
32611-8440}
\date{\today}

\maketitle

\begin{abstract}
Spin transport properties of a spin-polarized two-dimensional electron gas 
are studied in the presence of electron-electron interactions. 
Longitudinal and transverse spin diffusion coefficients are 
calculated with the quantum transport equation 
which includes many-particle effects in the random phase approximation. 
We find that the $e$-$e$ scattering, 
which does not contribute to 
charge drift mobility, has a significant contribution to the spin diffusion.
Thus, $e$-$e$ interaction 
causes non-negligible effects on the operations of proposed devices
dependent on spin transport in semiconductor heterostructures.
\end{abstract}

\pacs{72.25.Dc,72.80.Ey,73.40.-c,75.40.Gb}

\sloppy
\narrowtext
\section{Introduction}
Spin dependent electronic transport phenomena in semiconductor 
nanostructures are extensively studied in recent years, 
with motivations to develop novel electronic and optical devices 
based on carrier spin dynamics such as spin transistors \cite{SpinFET}, 
polarization dependent optical modulators, and even quantum computers. 
Recent experimental progress in transporting electronic spins
\cite{SpinTransport} over the distance
of 100 $\mu$m and injecting spins\cite{CarrierInjection} 
from ferromagnetic materials to compound semiconductors has been 
accelerating the prospect of these devices.

Charge transport is extensively studied to understand the 
operations of devices. 
In the conventional devices where the electron distribution is 
spatially homogeneous and the electrons move in the same direction 
irrespective of spin directions, the charge transport is expressed 
in terms of drift mobilities. When an electronic band is parabolic and 
umklapp processes are negligible, electron-electron ($e$-$e$) scatterings 
do not directly affect mobilities since the $e$-$e$ scattering conserves 
the total momentum of the system. \cite{Ben}
It influences the charge transport only indirectly by modifying
momentum distribution functions and screening other scattering processes of 
electrons.
Mobility is determined by $e$-phonon and $e$-ionized impurity scatterings. 
In high-quality, modulation-doped heterostructures of III-V semiconductors, 
electron mobility increases as temperature decreases and reaches 
the maximum value limited by the scattering with remote dopants.\cite{mobility}

In contrast to the charge transport, 
spin-up electrons move against spin-down electrons or vice versa in
the spin transport. 
Thus, in addition to $e$-ionized impurity scattering, 
we expect that
the Coulomb interaction between spin-up and -down electrons also contributes 
to the spin transport.\cite{Irene}
(The Coulomb interaction between two electrons with the same spin direction 
does not directly influence the spin transport as it conserves the spin current.)
As we show later, the $e$-$e$ interaction is as important as the 
$e$-ionized impurity scattering to the spin transport.

The spin transport is also influenced by the many-particle effects of
interacting electrons.
An electronic band dispersion is renormalized, giving the effective
mass of an electron somewhat different from the band effective mass,
and the bare Coulomb interaction is screened, leading to the 
$e$-$e$ interaction with a finite range.
These many-particle effects affect collision dynamics and modify 
transport coefficients.

The many-particle effects also have consequences on spin dynamics itself.
The spin axis of a noninteracting particle precesses around external
magnetic field $\bbox{\rm B}_{ex}$.
In the interacting system, each electron feels effective magnetic field
$\bbox{\rm B}_{mol}(\bbox{\rm r})$
due to the molecular field of surrounding electrons. The effective field
is related (almost proportional) 
to the local spin density and parallel to its direction 
as explicitly shown in our analysis below.
Thus in a spin polarized electron gas, a spin axis precesses around 
$\bbox{\rm B}_{ex} + \bbox{\rm B}_{mol}$.
This induces a ``spin-rotation term'' in the equation of motion for
the spin current, leading to a anomalous spin diffusion coefficients
and spin waves,\cite{spinrotation,spinsystem}
which were experimentally observed in $^3$He, $^3$He-$^4$He mixture,
and in atomic Hydrogen.\cite{experiment,3He}

In this paper we present a theory and numerical calculations of spin diffusion 
coefficients in a two-dimensional electron gas (2DEG) interacting through 
the Coulomb force. 
In our previous publication\cite{previous} 
we adopted the Hartree-Fock approximation (HFA)
to incorporate the many-particle effects in the lowest order.
Here we adopt the random phase approximation (RPA)
to improve the treatment on many-particle aspects of electrons
and obtain more reliable numerical results.
We compare the contributions from $e$-$e$ scattering and 
$e$-ionized impurity scattering to the spin diffusion in semiconductor 
heterostructures and discuss their relative importance.

This paper is organized as follows:
In Sec. II, our calculation scheme
is briefly described.
In Sec. III the numerical results for the spin transport coefficients
are shown with detailed discussions.
Conclusions are given in Sec. IV.
The differences in mathematical expressions from our previous calculation 
are summarized in Appendix.

\section{Theory}
Our theoretical model is already presented in our previous publication
\cite{previous} and we describe it only briefly here.
Several relevant expressions arising from the RPA will be
given in the appendix.
We consider two-dimensional electrons interacting through
the Coulomb interactions.
Other interactions such as electron-phonon, 
electron-ionized impurity or the exchange interaction of electrons with
magnetic impurities are not included in the present study.
We assume that electronic spins are polarized in the growth direction,
$z$-axis (normal to the 2D plane), in the quasi-thermal equilibrium,
and a direction of local spin polarization 
can be slightly tipped away from $z$-axis by applying r.f.
field or by external spin injection.
The total sheet density of spin-up and spin-down
electrons, $n(\bbox{\rm r},t)=n_+(\bbox{\rm r},t)+n_-(\bbox{\rm r},t)$,
is assumed to be spatially and temporally constant, and the spin density,
$M(\bbox{\rm r},t)=n_+(\bbox{\rm r},t)-n_-(\bbox{\rm r},t)$,
varies slowly on a hydrodynamic scale.\cite{scales}
The spin polarization is given by
$P=(n_+-n_-)/(n_++n_-)$.
Spin relaxation is not included in the present analysis.

A moderate external magnetic field, $\bbox{\rm B}_{ex}$, can be 
applied perpendicular to the 2D plane.
We assume that the field is so weak that it only affects 
spinor space, leaving electron orbital motion intact.
Otherwise the orbital motion is quantized into the Landau levels, and the
system turns into the quantized Hall state, to which our present
theory does not apply.
The estimation in our earlier work\cite{previous} 
shows that the {\em weak field
condition} is satisfied when $\bbox{\rm B}_{ex} < $ 1 T.
When this condition is satisfied,
the external field can trivially be erased from the theory by using
a rotating frame.

The quantum transport equation for the electron distribution functions
$\underline n_k$ is given by\cite{general,spinsystem}
\begin{eqnarray}
\label{quantumtransport}
{{\partial \underline n_p(\bbox{\rm r},t)} \over {\partial t}}+
{1 \over 2}\left\{ {{{\partial \underline \varepsilon _p(\bbox{\rm r},t)} 
\over {\partial \bbox{\rm p}}},{{\partial \underline n_p(\bbox{\rm r},t)} 
\over {\partial \bbox{\rm r}}}} \right\}-{1 \over 2}\left\{ 
{{{\partial \underline \varepsilon _p(\bbox{\rm r},t)} \over 
{\partial \bbox{\rm r}}},{{\partial \underline n_p(\bbox{\rm r},t)} \over 
{\partial \bbox{\rm p}}}} \right\}+{i \over \hbar }\left[ 
{\underline \varepsilon _p(\bbox{\rm r},t),
\underline n_p(\bbox{\rm r},t)} \right] \cr
=\left( {\partial \underline n_p(\bbox{\rm r},t)} 
\over {\partial t} \right )_{col.},
\end{eqnarray}
where the electron energy $\underline \varepsilon_k$ and the distribution
function $\underline n_k$ are 2$\times$2 matrices in spinor space, denoted
by underlining.
The diagonal components of $\underline n_k$ represent the population 
of spin-up and spin-down electrons and its off-diagonal elements represent 
the couplings between spin-up and spin-down states. 
The transport equation is similar to the Boltzmann equation except 
that there are anticommutators in the drift terms (the second and the
third term on the left) and a commutator in the last term on the left.
This term in the commutator is called the ``spin-rotation term'',
representing the effect of the external and molecular fields. 

The electron (quasiparticle) energy $\underline \varepsilon_k$ 
is composed of a kinetic energy term, a coupling to the external field 
and a self-energy caused by the presence of other electrons. 
In our earlier work,\cite{previous} 
we adopted the HFA to calculate an electron self-energy 
since the simple analytic expressions of the HFA considerably reduces
the load of numerical calculations. 
But the HFA does not describe the many-particle properties of 
2DEG appropriately: The energy dispersion of a quasiparticle 
is not given properly with a vanishing quasiparticle effective mass 
at Fermi surface due to the infinite range of the bare Coulomb interaction. 
Since the effective mass is one of the important parameters in our theory, 
a certain degree of arbitrariness exists in our previous calculations. 
And since the HFA does not include correlation terms, 
the self-energy for a spin-up electron does not contain the contribution 
from spin-down electrons. 
This leads us to an unrealistic situation when the spin polarization is high, 
i.e., provided that we prepared 2DEG with all spins pointing downward, 
and we put a spin-up electron into this 2DEG, 
then the energy dispersion of this spin-up electron remains that of 
a noninteracting electron in spite of the presence of many spin-down electrons. 
This prevented us from calculating the diffusion coefficients of 
highly polarized 2DEG in our earlier work.

In the present study we improve our calculation by adopting the RPA.
It takes into account exchange and correlation (though not completely), 
and we can correct the problems of the HFA mentioned above. 
The RPA self-energy for an electron gas is discussed in many text books 
(e.g., see Ref.\onlinecite{Mahan} and \onlinecite{HaugKoch}), 
and its detailed expressions
for a two-dimensional, multi-band electron system are given by
Vinter\cite{Vinter} and DasSarma {\em et. al}.\cite{DasSarma} 
We use the standard RPA self-energy without vertex corrections.
The Matsubara function (imaginary time Green's function) for
the self-energy, which is a 2$\times$2 matrix in spinor space, 
is written as
\begin{equation}
\label{RPAself-energy}
\hbar \Sigma _{\alpha ,\beta }^{RPA}(\bbox{\rm k},ik_m)=
-{1 \over {\beta V}}\sum\limits_{\bbox{\rm q}} {\sum\limits_{iq_n} 
{V_{\alpha ,\beta }^{RPA}(-\bbox{\rm q},-iq_n)}
G_{\alpha ,\beta }^0(\bbox{\rm k}+\bbox{\rm q},ik_m+iq_n)},$$
\end{equation}
where $\alpha$ and $\beta$ are indices for spin,
$G_{\alpha ,\beta }^0$ is the noninteracting electron propagator.
($G_{\alpha ,\beta }^0$ is finite only when $\alpha = \beta$.)
The RPA screened Coulomb interaction $V_{\alpha ,\beta }^{RPA}$
satisfies the Dyson equation,
\begin{equation}
\label{screenedCoulomb}
V_{\alpha ,\beta }^{RPA}(\bbox{\rm q},iq_n)=
V_{\alpha ,\beta }(q)+\sum\limits_{\delta ,\delta '} {V_{\alpha ,\delta }(q)
P_{\delta ,\delta '}^0(\bbox{\rm q},iq_n)
V_{\delta ,\beta }^{RPA}(\bbox{\rm q},iq_n)},
\end{equation}
where $V_{\alpha ,\beta }(q)$ is the bare Coulomb interaction.
The lowest order polarization (bubble diagram) is finite only when 
$\delta = \delta '$, given by
\begin{equation}
\label{polarization}
P_{\delta ,\delta }^0(\bbox{\rm q},iq_n)=
{1 \over {\beta V}}\sum\limits_{\bbox{\rm p}} {\sum\limits_{ip_m} 
{G_{\delta ,\delta }^0(\bbox{\rm p},ip_m)}
G_{\delta ,\delta }^0(\bbox{\rm p}+\bbox{\rm q},ip_m+iq_n)}.
\end{equation}
The real-time functions are obtained by the analytic continuation
after the momentum integration and the summation over Matsubara
frequencies.
The quasiparticle energy dispersion is obtained from the 
real part of the self-energy 
following the prescription suggested by Rice.\cite{Rice}

In Fig. 1 we show energy-momentum dispersions of spin-up and -down 
electrons in 2DEG with spin-polarizations 
$P =$ 0.1 -- 0.9 
and with total electron sheet density at 2 $\times$ $10^{11}$ cm $^{-2}$. 
The electron energy is measured relative to 
the zero of the noninteracting electron dispersion. 
One of the important many-particle effects in the interacting electrons 
is the reduction of one-electron energy (Bandgap renormalization). 
Since $n_+ > n_-$, the renormalization is larger in the spin-up electron. 
The energy differences between the spin-up and 
the spin-down electron increase with spin polarizations. 
When $P <$ 0.5, the energy dispersions of both spin-up and -down electrons 
are well approximated by parabolic band dispersions. 
We obtained the quasiparticle effective mass $m^*$ between 1.11$m$ and 
1.19$m$ depending on spin polarizations. 
($m$ is a band effective mass. We used $m = 0.067m_0$ for 
the conduction band of GaAs.) 
In numerical computations of transport coefficients,
we have used the parabolic band approximation for the 
energy dispersions.

The quantum transport equation eq.(\ref{quantumtransport})
is solved by the Chapman-Enskog expansion 
to the lowest order using variational solutions.\cite{McLennan}
The final expression of the equation for the spin current
is obtained by multiplying the 
transport equation by the particle velocity and integrating
over the momentum as,\cite{previous}
\begin{eqnarray}
\label{equationforJ}
{{\partial J_{\sigma ,i}(\bbox{\rm r},t)} \over {\partial t}}+
\bbox{\hat e}_\sigma{{\partial M(\bbox{\rm r},t)} \over {\partial r_i}}
A\left( {k_{F\pm }} \right)+
M(\bbox{\rm r},t){{\partial \bbox{\hat e}_\sigma} \over 
{\partial r_i}}B\left( {k_{F\pm }} \right)
+\mu_{SR}\bbox{\rm M}\times J_{\bbox{\sigma} ,i}
+\gamma_{gy}\bbox{\rm B}_{ex}\times J_{\bbox{\sigma} ,i} \cr
=-{1 \over {\tau _{\parallel}}}J_{\bbox{\sigma}_\parallel,j}
-{1 \over {\tau _\bot }}J_{\bbox{\sigma} _\bot ,j},
\end{eqnarray}
where $J_{\bbox{\rm \sigma},j}$ is a spin current representing the 
flow of the spin component $\bbox{\rm \sigma}$ in the spatial
direction $j = x, y, z$, 
$\bbox{\hat e}_{\sigma}$ is a unit vector in the direction of spin polarization,
and $\bbox{\rm M} = M\bbox{\hat e}_{\sigma}$.
The factor $\mu_{SR}$, which is a function of $k_{F\pm}$,
represents the contribution of the molecular field, and
$\gamma_{gy} = -g\mu_B/\hbar$ is a gyromagnetic ratio.
The coefficients $A\left( {k_{F\pm }} \right)$, $B\left( {k_{F\pm }} \right)$
and $\mu_{SR}$ will be given in the appendix.

\section{Results and Discussion}
\subsection{Spin diffusion coefficients}
We show the results of numerical calculations for 
spin diffusion coefficients.
Since a spin current is a tensor defined by the two vectors; 
the direction of spatial flow and the direction of spin, 
we should distinguish longitudinal ($D_{\parallel}$) and transverse 
($D_{\perp}$) spin diffusion coefficients. (See Fig. 2) 
In the longitudinal spin diffusion, consider the system where the direction 
of spin polarization is aligned in $z$ everywhere but its magnitude has 
gradient in the direction $x$. 
Then the longitudinal spin current flows in the spatial direction $x$ 
whose spin direction is in $z$, parallel to the spin polarization. 
While in the transverse spin diffusion, consider the system where the 
magnitude of spin polarization is the same everywhere 
but its direction is tilted gradually from $z$ to $x$ by a small angle 
when we go from $x = x_0$ to $x_0 + dx$. 
In this case the transverse spin current flows in 
the spatial direction $x$ with the spin direction in $x$, 
perpendicular to the spin polarization. 
These two diffusions show different magnitudes and temperature 
dependence.\cite{Meyerovich}

The temperature dependence of $D_{\parallel}$ and $D_{\perp}$ 
is plotted in Fig. 3 and 4, respectively, with 0.01 $< P <$ 0.5. 
We calculate in the degenerate region, $T <$ 20 K. 
The Fermi temperature is 83 K in the unpolarized 2DEG 
with the sheet density 2 $\times$ 10$^{11}$ cm$^{-2}$
(Fermi wave number $k_F$ = 0.112 nm$^{-1}$). 
The magnitudes of the diffusion coefficients are somewhat larger 
than those calculated previously in Ref.\onlinecite{previous} 
but their temperature dependencies are identical. 
We believe that the present results calculated with the RPA should be 
more accurate in predicting the transport coefficients. 
As shown in Fig. 3, $D_{\parallel}$ increases at low temperatures 
and diverges as $T^{-2}$ when $T$ approaches zero. 
A close inspection shows it deviates slightly from $T^{-2}$ dependence 
(bulging downward). 
Actually we find $D_{\parallel} \propto (E_F/k_BT)^2/{\rm ln}(E_F/k_BT)$ 
as expected of the ``normal'' transport coefficient in 2D degenerate Fermions. 
(The logarithmic correction factor is characteristic of a 2D system.
\cite{logcorrection}) 
While in Fig. 4, $D_{\perp}$ increases in the higher temperatures in 
a similar manner as $D_{\parallel}$, but it departs from $T^{-2}$ 
dependence and becomes constant as $T \to 0$ at the larger $P$.
 
The different temperature dependence of $D_{\parallel}$ and 
$D_{\perp}$ is attributed to the phase space available in 
the collisions of two diffusion processes. 
In the longitudinal spin diffusion, a close inspection of the associated 
collision term shows that electrons in the vicinity of the Fermi surface 
within the width $k_BT$ only are allowed to participate in the collision 
due to the kinematics and the Pauli exclusion. 
Thus as the temperature is lowered, the number of electrons participate 
in the collision decreases, leading to the reduction of a collision rate. 
While in the transverse case, all electrons between the two Fermi surfaces 
of the spin-up and -down electrons participate in the collision. 
Thus the larger the spin polarization is, 
the more electrons are involved in the collision, 
leading to the finite and constant diffusion coefficients at $T \to 0$. 
This anomalous temperature dependence is observed experimentally in 
$^3$He-$^4$He system.\cite{3He}

\subsection{Electron-impurity scattering}
In the previous subsection we have discussed the spin diffusion 
of 2DEG in the presence of $e$-$e$ scattering alone. 
The spin diffusion in actual samples of heterostructures and
quantum wells (QWs) should also be affected by other scattering processes.
In heterostructures of III-V semiconductors, a low temperature electron 
{\em mobility}, $\mu_e$, is dominated by the scattering with 
remote ionized impurities (dopants). 
We expect that the $e$-impurity scattering will also strongly 
influence the spin transport. 
We compare the spin diffusion coefficients limited by $e$-$e$ 
scattering which we have calculated above
with those limited by $e$-impurity scattering,
and show that the $e$-$e$ scattering bears substantial contribution to
the spin transport in actual sample structures,
although the $e$-$e$ scattering has only an indirect consequence to the 
{\em charge} transport.
We do not give the explicit calculations of the spin diffusion
including the $e$-impurity scattering but we estimate its magnitude from 
experimental carrier mobilities.

The highest mobility reported so far in GaAs/AlGaAs heterostructures
exceeds 1 $\times$ 10$^7$ cm$^2$ V$^{-1}$ s$^{-1}$ below 1 K.\cite{Pfeiffer} 
The {\em charge} diffusion coefficient $D_e$ arising from
$e$-impurity scattering is related to the mobility $\mu_e$
through the generalized Einstein's relation.\cite{Ashcroft}
For a degenerate system, we have
$${{eD_e} \over {\mu _ek_BT}}=(1+e^{-\mu /k_BT})\log (1+e^{\mu /k_BT}),$$
where $\mu$ is a chemical potential.
In the high quality samples cited above, the charge diffusion coefficient 
$D_e$ is estimated to be 9 $\times$ 10$^4$ cm$^2$ s$^{-1}$ below 10 K. 
Assuming that the magnitude of the {\em spin} diffusion limited by 
$e$-impurity 
scattering is similar to the charge diffusion, we obtain the estimated  
value of 9 $\times$ 10$^4$ cm$^2$ s$^{-1}$ for the spin diffusion,
which is, compared with Fig. 3, close to the spin diffusion
limited by $e$-$e$ scattering at the lowest temperatures.
Thus the $e$-$e$ scattering has a significant
contribution to the spin diffusion, comparable to the $e$-impurity
scattering. 
In moderate-quality samples with 
$\mu \le$ 1 $\times$ 10$^6$ cm$^2$ V$^{-1}$ s$^{-1}$, 
the relative weight of $e$-ionized impurity scattering should 
become larger than that of $e$-$e$ scattering.

\subsection{Effective magnetic field}
In a many-electron system an individual electron interacts with surrounding 
electrons through the Coulomb interaction.
This effect is represented in the Fermi liquid theory 
that an electron is subject to
a molecular (mean) field produced by surrounding electrons.
Due to the exchange part of the interaction, the energy of an 
individual electron depends on the spin directions of itself and neighboring
electrons.
Thus the molecular field acts as an effective magnetic field 
$\bbox{\rm B}_{mol}$
to the individual electron, and its spin precesses around
$\bbox{\rm B}_{mol}$.
The consequence of $\bbox{\rm B}_{mol}$ was first pointed out by Leggett and 
Rice\cite{spinrotation,BaymPethick}
that the effective transverse spin diffusion coefficients 
measured by spin echo experiments should depend on the spin-tipping 
angle by the initial r.f. pulse (Leggett-Rice effect).
The anomalous behavior of the transverse 
spin diffusion and the spin waves discussed in
our previous publication\cite{previous} 
have its physical origin in the precession of flowing spins around 
the effective magnetic field.
Consequences of the effective magnetic field can be seen 
explicitly in
the last two terms on the left-hand side of eq. (\ref{equationforJ}),
$\mu_{SR}\bbox{\rm M}\times J_{\bbox{\sigma} ,i}
+\gamma_{gy}\bbox{\rm B}_{ex}\times J_{\bbox{\sigma} ,i}$.
The second term expresses the spin precession around 
the external magnetic field,
while the first term represents the spin precession 
around the effective magnetic
field. The effective field is parallel to 
spin density $\bbox{\rm M}$, and its magnitude depends on 
$\mu_{SR}$ and $M$.
We should notice that the precession occurs not on the spin density
but on the spin current $J_{\bbox{\sigma}}$, {\em i.e.},
the precession occurs on an electron flowing into the region
with its spin direction slightly tipped away from 
the local spin density.
(If the direction of $\bbox{\rm M}$ is constant throughout the system,
the effective field $\bbox{\rm B}_{mol}$ has no net effect.)
In Fig. 5 we plot the magnitude of the effective field
in terms of spin splitting energy, 
$\Delta E_{mol}=\hbar \mu_{SR} M$ [eV],
(or the Larmor precession frequency) as a 
function of spin polarization $P = M/n$.
The spin splitting shows a sublinear dependence on the spin polarization
since $\mu_{SR}$ decreases with $P$.

\section{Conclusions}
We have calculated spin diffusion coefficients of 2DES 
interacting through the Coulomb force.
The improved approximation in the RPA leads to a more precise
treatment of many-particle aspects of the electron properties
compared with our earlier calculations using the HFA.
The magnitudes of both the longitudinal and the transverse 
spin diffusion coefficients are larger than our previous calculations
but their temperature dependencies are qualitatively 
similar to the previous ones.

Charge transport in heterostructures, at low temperatures, 
is dominated by the electron-ionized impurity scattering,
while the $e$-$e$ scattering has little consequences.
In contrast, as we have shown, the spin transport is
largely affected by the $e$-$e$ scattering in high-quality samples.
Thus we expect that the spin transport properties in semiconductor
devices should depend on spin-up and -down carrier distributions 
both in real and momentum space, in addition to the distributions of
remote dopants.
This fact should be taken into account
in designing recently proposed spintronics devices.

\acknowledgments
This work is supported 
by a Grant-in-Aid for Scientific Research from the Ministry of 
Education, Science, Sports and Culture of Japan.

\appendix
\section{Expressions in RPA}
The equations are shown in detail in our previous publication
\cite{previous} calculated in the HFA.
In the present study, the mathematical expressions can be
obtained by substituting the self-energy terms calculated
in the RPA for those in the HFA.
In contrast to the HFA self-energy, which gives an analytic expression,
the RPA expressions require numerical integrations.
We give the relevant expressions used in the present study below.

The quasiparticle energy $\underline \varepsilon _p^0(\bbox{\rm r},t)$
in the local equilibrium (Eq. (2.4) in 
Ref. \onlinecite{previous}) is obtained from
the real part of the RPA self-energy 
following the prescription suggested by Rice,\cite{Rice}
\begin{equation}
\label{quasienergy}
\varepsilon _{p\pm}^0(\bbox{\rm r},t) 
= \varepsilon_{Kin}+
{\rm Re} \hbar \Sigma _{\pm}^{RPA}
(\bbox{\rm p},\hbar\omega=\varepsilon_{Kin};\bbox{\rm r},t),
\end{equation}
where $+$ ($-$) for (1,1) ((2,2)) component, respectively, and the 
off-diagonal elements are zero.
$\varepsilon_{Kin} = p^2 / 2m$ is the noninteracting energy of an electron,
and $\hbar \Sigma _{\pm}^{RPA}(\bbox{\rm p},\hbar\omega;\bbox{\rm r},t)$ 
is obtained from the 
imaginary-time function in Eq. (\ref{RPAself-energy}).
In the equation of motion for spin current $J_{\bbox{\rm \sigma},j}$,
the coefficient $A$ is a function of spin-up and spin-down
electron densities (or the Fermi wave numbers) and is given by
\begin{equation}
\label{A}
A\left( {k_{F\pm },\bbox{\rm r},t} \right)=
{1 \over2 m}\left( {{{n_+(\bbox{\rm r},t)} \over {\bar G_+}}+
{{n_-(\bbox{\rm r},t)} \over {\bar G_-}}} \right),
\end{equation}
where $\bar G_\pm ^{-1}=G_\pm ^{-1}-\alpha _\pm $ with
$$G_\pm ={{m_\pm ^*} \over {2\pi \hbar ^2}}, \ \
\alpha _\pm ={{\partial \varepsilon^0 _{p=\hbar k_{F\pm },\pm }} 
\over {\partial n_+}}-{{\partial 
\varepsilon^0 _{p=\hbar k_{F\pm },\pm }} \over {\partial n_-}}.$$
The coefficient $B$ is
\begin{equation}
\label{B}
B\left( {k_{F\pm },\bbox{\rm r},t} \right)=
{{\pi \hbar ^2n} \over {mm^*}} 
+{1 \over m}\left( {{{\Delta \varepsilon _{p=\hbar k_{F+}}n_+} 
\over {n_+-n_-}}+{{\Delta \varepsilon _{p=\hbar k_{F-}}n_-} 
\over {n_+-n_-}}} \right),
\end{equation}
where $\Delta \varepsilon _{p}$ is defined as,
$$\Delta \varepsilon _{p}(\bbox{\rm r},t)
={1\over 2}\left( \varepsilon _{p+}^0(\bbox{\rm r},t)
-\varepsilon _{p-}^0(\bbox{\rm r},t)\right).$$ 
The spin-rotation parameter $\mu_{SR}$ is obtained by the numerically 
integration;
\begin{equation}
\label{musr}
\mu _{SR}=-{1 \over {\pi \hbar n M^2}}
\int {{{d^2k} \over {(2\pi )^2}}}
k^2\left( {n_{k+}^0-n_{k-}^0} \right)
\Delta \varepsilon _{k}.
\end{equation}
The collision terms are given by the same expressions as given in Ref.
\onlinecite{previous} except that
the quasiparticle energies in the RPA are used.

\begin{figure}
\caption{The inplane energy dispersions of interacting, 
spin-polarized 2D electrons with a total sheet density 2 $\times$ 
10$^{11}$ cm$^{-2}$ calculated in the RPA. 
The energy is measured relative to 
the zero of a noninteracting electron dispersion. 
The Fermi surfaces for spin-up (spin-down) 
electrons are at 0.1127 nm$^{-1}$ 
(0.1115 nm$^{-1}$) for $P$ = 0.1, 0.1373 nm$^{-1}$ (0.0793 nm$^{-1}$) 
for $P$ = 0.5, and 0.1545 nm$^{-1}$ (0.0354 nm$^{-1}$) 
for $P$ = 0.9, respectively.
}
\end{figure}

\begin{figure}
\caption{The definitions of a longitudinal spin diffusion coefficient 
$D_{\parallel}$ and a transverse spin diffusion coefficient 
$D_{\perp}$. See the text for details.} 
\end{figure}

\begin{figure}
\caption{The temperature dependence of the longitudinal spin diffusion 
coefficients $D_{\parallel}$ with 0.01 $< P <$ 0.5. 
The total electron sheet density is fixed at 2 $\times$ 10$^{11}$ cm$^{-2}$.}
\end{figure}

\begin{figure}
\caption{The temperature dependence of the transverse spin diffusion 
coefficients $D_{\perp}$ with 0.01 $< P <$ 0.5. 
The total electron sheet density is fixed at 2 $\times$ 10$^{11}$ cm$^{-2}$.}
\end{figure}

\begin{figure}
\caption{
The magnitude of the effective magnetic field due to the molecular field
expressed in terms of the spin splitting energy (left) or the Larmor
precession frequency (right).}
\end{figure}

\centerline{\epsfbox{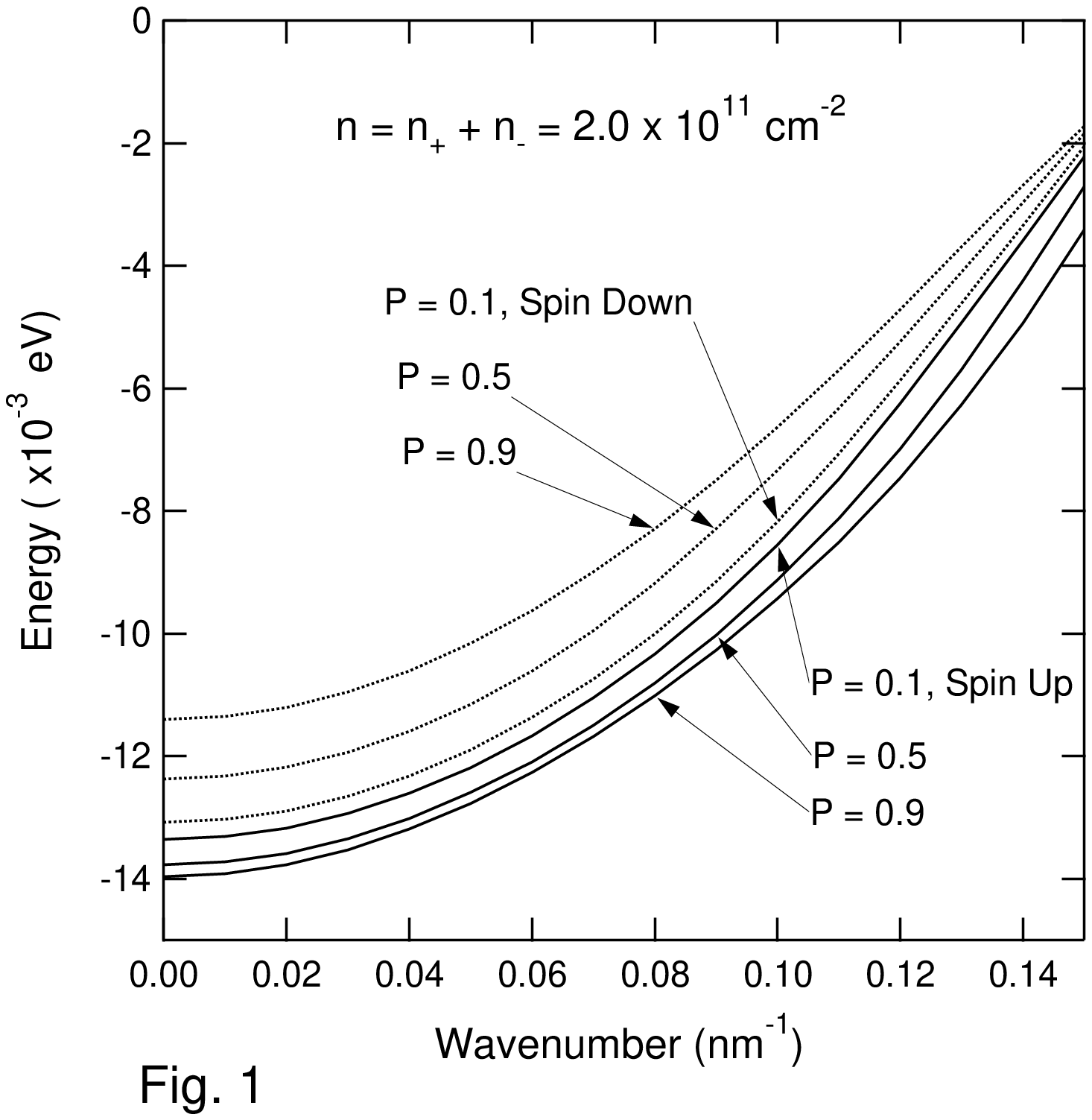}}
\centerline{\epsfbox{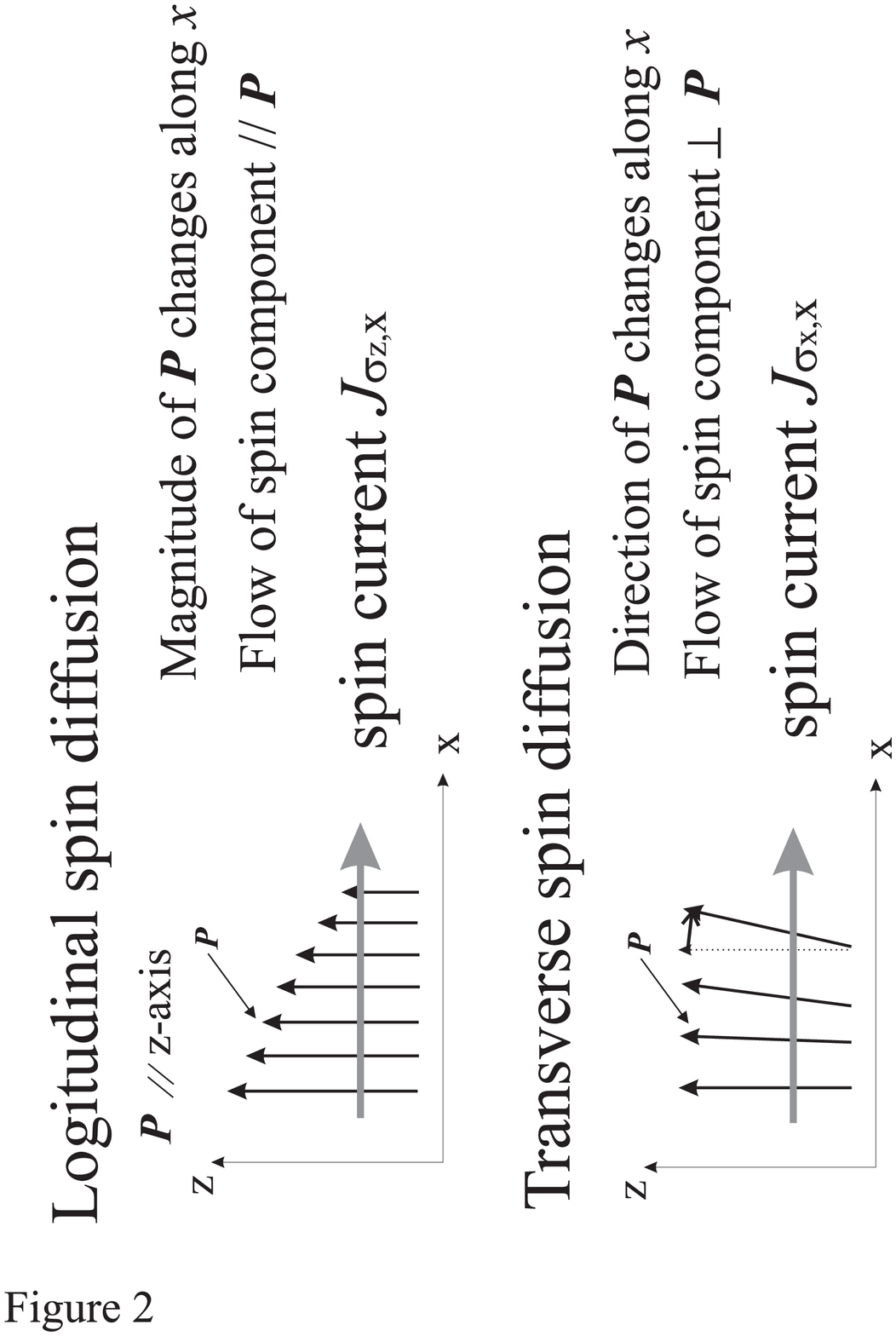}}
\centerline{\epsfbox{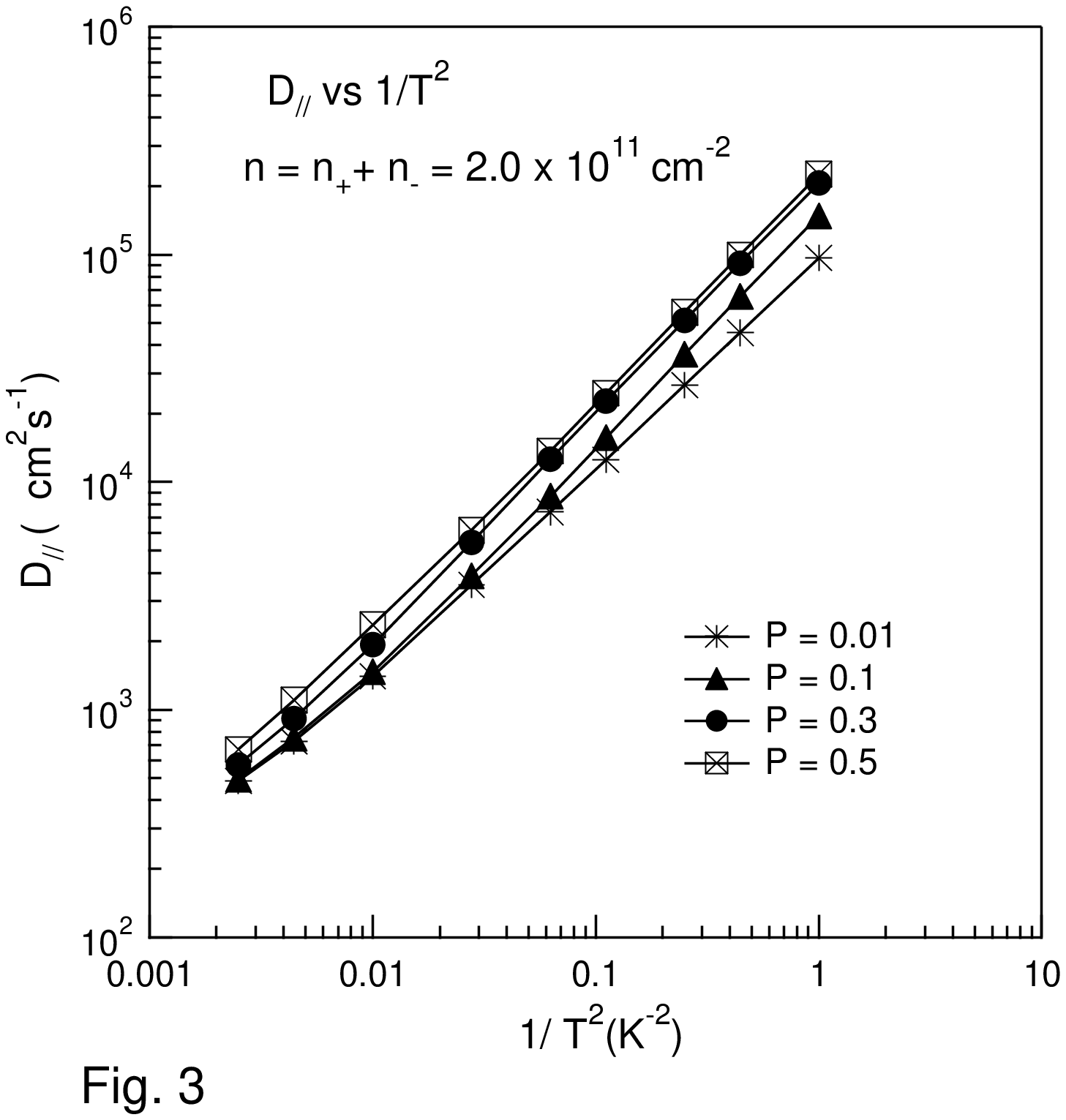}}
\centerline{\epsfbox{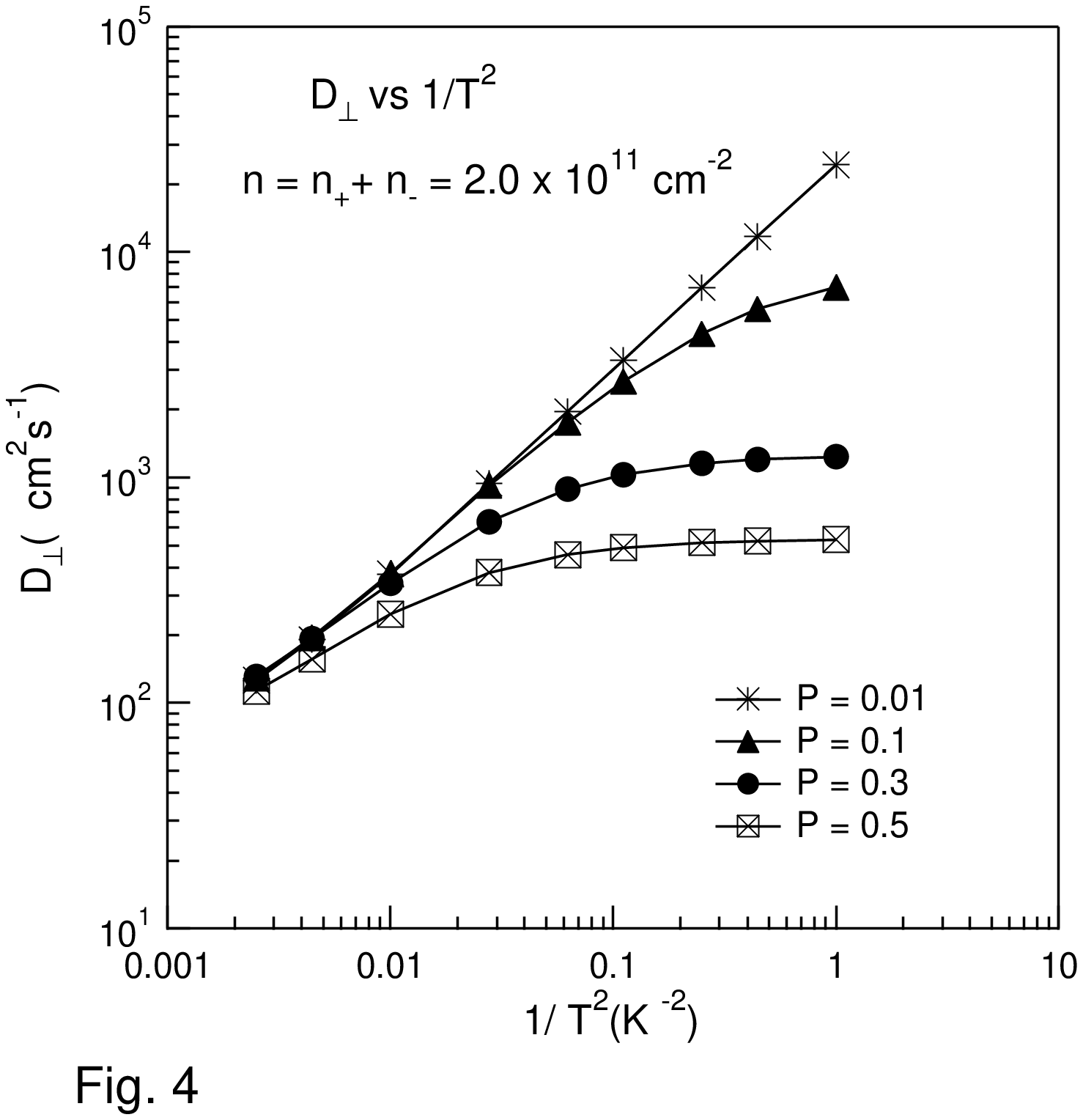}}
\centerline{\epsfbox{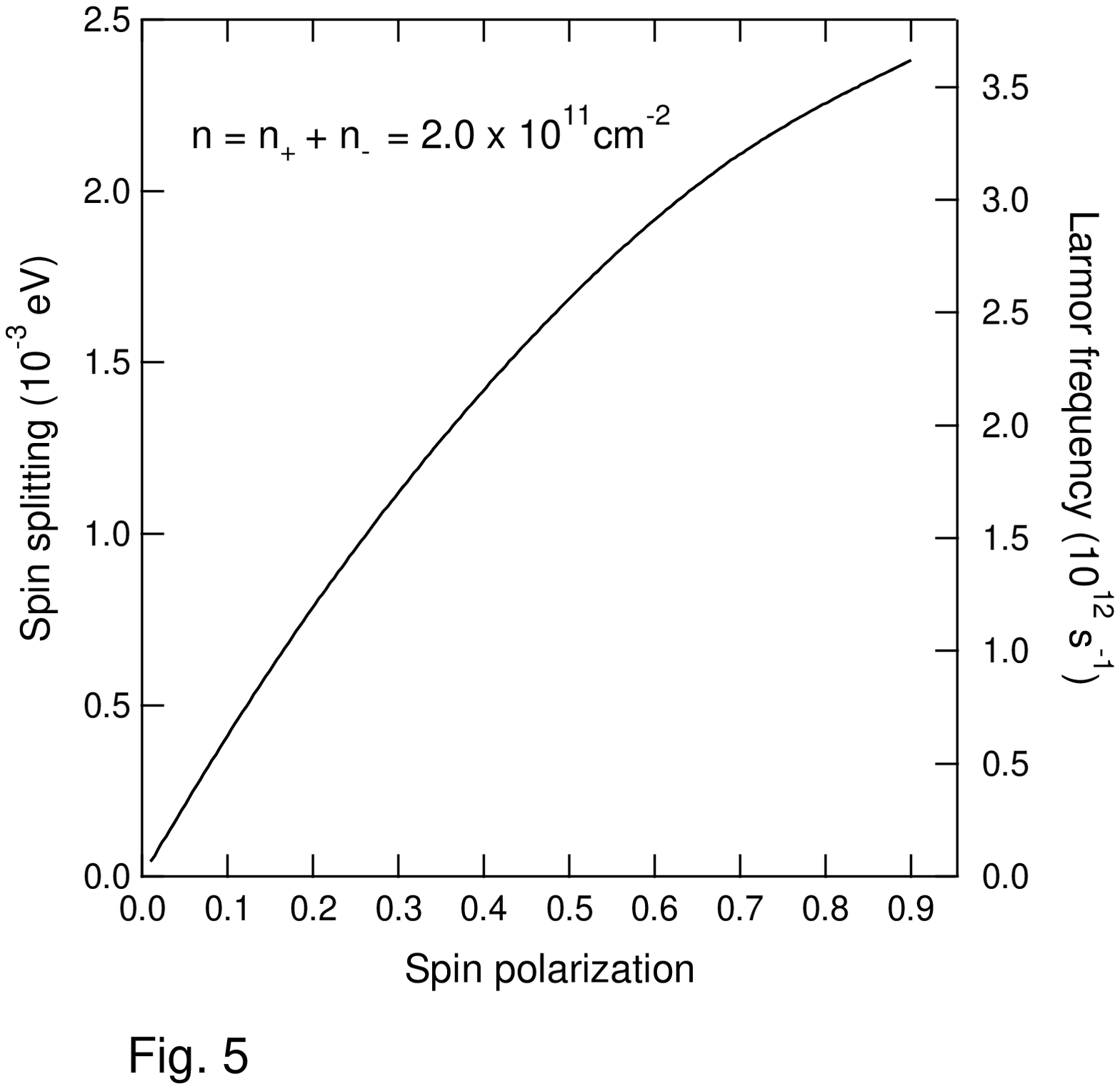}}


\begin{references}

\bibitem[*]{email}
E-mail Address: takahasy@eie.yz.yamagata-u.ac.jp

\bibitem{SpinFET}
S.~Datta and B.~Das,
Appl. Phys. Lett. {\bf56}, 665 (1990).

\bibitem{SpinTransport}
J. M. Kikkawa and D. D. Awschalom,
Nature {\bf397}, 139 (1999).

\bibitem{CarrierInjection}
R. Fiederling, M. Keim, G.Reuscher, W. Ossau, G. Schmidt, A. Waag,
and L. W. Molenkamp, Nature {\bf402}, 787 (1999);
Y. Ohno, D. K. Young, B. Beschoten, F. Matsukura, H. Ohno,
and D. D. Awschalom, {\em ibid}. {\bf402}, 790 (1999).

\bibitem{Ben}
Ben Yu-Kuang Hu and Karsten Flensberg,
Phys. Rev. B {\bf33}, 10072 (1996).

\bibitem{mobility}
K. Hirakawa and H. Sakaki, 
Phys. Rev. B {\bf33}, 8291 (1986); 
T. Saku. Y. Horikoshi and Y. Tokura, 
Jpn. J. Appl. Phys. {\bf35}, 34 (1996).

\bibitem{Irene}
During the preparation of this manuscript we find that
this point is also discussed in 
Irene D'Amico and Giovanni Vignale,
Phys. Rev. B {\bf 62}, 4853 (2000).

\bibitem{spinrotation}
A.~J.~Leggett and M.~J.~Rice,
Phys. Rev. Lett. {\bf20}, 586 (1968);
A.~J.~Leggett,
J. Phys. C {\bf3}, 448 (1970).

\bibitem{spinsystem}
J.~W.~Jeon and W.~J.~Mullin,
J. Phys. (Paris) {\bf49}, 1691 (1988);
A.~E.~Ruckenstein and L.~P.~L\'evy,
Phys. Rev. B {\bf39}, 183 (1989);
W.~J.~Mullin and J.~W.~Jeon,
J. Low Temp. Phys. {\bf88}, 433 (1992).

\bibitem{experiment}
N.~Masuhara, D.~Candela, D.~O.~Edwards, R.~F.~Hoyt, H.~N.~Scholz,
D.~S.~Sherrill, and R.~Combescot,
Phys. Rev. Lett. {\bf53}, 1168 (1984);
J. R. Owers-Bradley, H. Chocholacs, R. M. Mueller, Ch. Buchal,
M. Kubota, and F. Pobell,
{\em ibid}. {\bf51}, 2120 (1983); 
L. R. Corruccini, D. D. Osheroff, D. M. Lee, and R. C. Richardson,
{\em ibid}. {\bf27}, 650 (1971);
J. Low Temp. Phys. {\bf8}, 229 (1972).

\bibitem{3He}
L-J. Wei, N. Kalechofsky and D. Candela, 
Phys. Rev. Lett. {\bf 71}, 879 (1993); 
J. H. Ager, R. M. Bowley, R. Konig, and J. R. Owers-Bradley, 
Phys. Rev. B {\bf 50}, 13062 (1994).

\bibitem{previous}
Yutaka Takahashi, Kosuke Shizume, and Naoto Masuhara,
Phys. Rev. B {\bf 60}, 4856 (1999).

\bibitem{scales}
In the present study, a typical
hydrodynamic scale is few micrometers if we assume optical
excitations. A microscopic scale is $\le$ 100 nm determined by 
i) the interaction range,
ii) the interparticle spacings,
iii) the mean free path
of electrons.

\bibitem{general}
For general references to quantum transport, see
L.~P.~Kadanoff and G.~Baym,
{\em Quantum Statistical Mechanics} (Benjamin, New York, 1962);
D.~C.~Langreth and J.~W.~Wilkins,
Phys. Rev. B {\bf6}, 3189 (1972);
H.~Haug and A.-P.~Jauho,
{\em Quantum Kinetics in Transport and Optics of Semiconductors} 
(Springer, Berlin, 1996).

\bibitem{Mahan}
G.~D.~Mahan,
{\em Many-Particle Physics} 2nd ed. (Plenum, New York, 1990).

\bibitem{HaugKoch}
H. Haug and S. Koch,
{\em Quantum Theory of the Optical and Electronic Properties of Semiconductors} 
(World Scientific, Singapore, 1990).

\bibitem{Vinter}
B. Vinter,
Phys. Rev. B {\bf 13}, 4447 (1976).

\bibitem{DasSarma}
S. DasSarma, R. K. Kalia, M. Nakayama, and J. J. Quinn,
Phys. Rev. B {\bf 19}, 6397 (1979).
 
\bibitem{Rice}
T. M. Rice,
Ann. Phys. {\bf 31}, 100 (1965).

\bibitem{McLennan}
J. A. McLennan,
{\em Introduction to Nonequilibrium Statistical Mechanics}
(Prentice Hall, New Jersey, 1989).

\bibitem{Meyerovich}
A.~E.~Meyerovich,
Phys. Lett. {\bf107A}, 177 (1985).

\bibitem{logcorrection}
C. Hodges, H. Smith, and J. W. Wilkins,
Phys. Rev. B {\bf 4}, 302 (1971);
A. V. Chaplik,
Zh. Eksp. Teor. Fiz. {\bf60}, 1845 (1971) [Sov. Phys. JETP {\bf33}, 997 (1971)]. 

\bibitem{Pfeiffer}
L. Pfeiffer, K. W. West, H. L. Stormer and K. W. Baldwin, 
Appl. Phys. Lett. {\bf 55}, 1888 (1989); 
V. Umansky, R. de-Picciotto and M. Heiblum, 
{\em ibid}. {\bf 71}, 683 (1997).

\bibitem{Ashcroft}
See ch.29 of N. W. Ashcroft and N. D. Mermin,
{\em Solid State Physics}
(HRW, Philadelphia, 1976).

\bibitem{BaymPethick}
G.~Baym and C.~Pethick,
{\em Landau Fermi-liquid theory} (Wiley, New York, 1991).

\end{references}
\end{document}